\documentclass[12pt,preprint]{aastex}
\shortauthors{R. Subrahmanyan, A. J. Beasley, W. M. Goss, K. Golap, 
	\& R. W. Hunstead}
\shorttitle{Relic in a poor cluster}
\begin{document}

\title{PKS~B1400$-$33: an unusual radio relic in a poor cluster}

\author{Ravi Subrahmanyan}
\affil{Australia Telescope National Facility, CSIRO, 
Locked bag 194, Narrabri, NSW 2390, Australia}

\author{A. J. Beasley}
\affil{Owens Valley Radio Observatory,
	California Institute of Technology,
	Pasadena, CA 91125}

\author{W. M. Goss and K. Golap}
\affil{National Radio Astronomy Observatory, PO Box 0,
		Socorro, NM 87801}

\author{R. W. Hunstead}
\affil{School of Physics, 
	University of Sydney, NSW 2006, Australia}

\begin{abstract}
We present new arcminute resolution radio images of the
low surface brightness radio source 
PKS~B1400$-$33 that is located in the poor cluster Abell S753. 
The observations consist of 330 MHz VLA,
843 MHz MOST and 1398 and 2378 MHz ATCA data.
These new images, with higher surface brightness sensitivity
than previous observations, reveal that
the large scale structure consists of extended filamentary emission 
bounded by edge-brightened rims. The
source is offset on one side of symmetrically distributed 
X-ray emission that is 
centered on the dominant cluster galaxy NGC 5419. 
PKS~B1400$-$33 is a rare example of a relic in a poor cluster with radio 
properties unlike those of most relics and halos observed in cluster
environments.

The diffuse source appears to have had an unusual origin and we discuss 
possible mechanisms.  We examine whether the source could be 
re-energized relic radio plasma
or a buoyant synchrotron bubble that is a relic of activity 
in NGC 5419. The more exciting prospect is that the source is
relic plasma preserved in the cluster gaseous environment following
the chance injection of a radio lobe into the ICM as a result of
activity in a galaxy at the periphery of the cluster.  
\end{abstract}

\keywords{galaxies: clusters --- radio continuum: galaxies --- 
	radio sources: general, diffuse}	

\section{Introduction}

Extended emission structures with steep radio spectral indices, 
which are not obviously associated with any optical galaxy, are sometimes
observed in rich cluster environments.  These sources 
are broadly separated into
cluster wide `halos' that appear relaxed and symmetric with respect to the 
X-ray intracluster gas, and the peripheral arc-like structures referred
to as `relics'.   Both types are predominantly observed to be
associated with hot ($T_{X} \ge 6$ keV),
X-ray luminous ($L_{X} > 4\times 10^{44}$ erg s$^{-1}$) rich 
compact clusters \citep{schuecker99}; however, the clusters
that have peripheral relics may have somewhat lower 
temperatures \citep{feretti99}.  A standard hypothesis is that these sources
are relic synchrotron plasma that has been revived.

PKS~B1400$-$33 is an extended radio source, with an 85 MHz flux 
density of 57 Jy \citep{mills60}, and has the lowest surface brightness of
any source in the Parkes catalog.  
PKS~B1400$-$33 appears to be associated with the poor cluster S753 of Abell
richness class 0; however, the source not been conclusively identified
with any optical galaxy.  Previous low frequency images
of the source using the VLA and the MOST \citep{goss87} showed 
a relatively bright rim 
along the NE edge and low surface brightness filamentary emission 
trailing off to the SW.    The source was observed
to have a steep spectrum with spectral index $\alpha$
($S_{\nu} \propto \nu^{-\alpha}$) in the range 1.2-2.4.

The unusual nature of this possible relic radio
source and its potential value as a probe of the gas dynamics and
evolution of poor cluster environments has led us to carry out new radio
observations of the source.  We have made images
with improved surface brightness sensitivity at 330 MHz using the Very Large
Array (VLA), at 843 MHz with the Molonglo Observatory Synthesis Telescope 
(MOST) and at 1398 and 2378 MHz
using the Australia Telescope Compact Array (ATCA).  Table 1 is a summary
of the new radio observations.  These are presented here followed by
discussions on the origin of this source in the light of the current 
understanding of the phenomenology and formation of relics in the 
intracluster medium (ICM).

\section{The radio continuum images}

PKS~B1400$-$33 was observed at 330 MHz with the VLA separately in the
CnB, C, and DnC array configurations in order to image
the extended emission in this 
low surface brightness source.  Hybrid arrays were included because of 
the southern declination.  The data were self calibrated and 
imaged in {\sc aips} using three-dimensional imaging routines.
The VLA 330 MHz image of the source 
is shown in Fig.~1.  The image has been made combining data
from all the configurations with a beam of FWHM
$77 \times 53$ arcsec$^{2}$ at a P.A. of $-24^{\circ}$; the
r.m.s. noise is 2.5 mJy beam$^{-1}$.  This radio image, as well as 
all the higher frequency radio images presented here, has been
corrected for the primary beam attenuation.

Because of its steep spectrum,
the total extent of the diffuse source is best defined by the 330 MHz
VLA image. The low surface brightness source is bounded to the
NE by a relatively bright rim of emission (marked NE rim in Fig. 1), 
while the source is
bounded on the opposite SW side by a second fainter ridge
of emission (marked SW rim in Fig. 1).  
Both the bounding rims are concave outwards.
The emission that lies between the bounding ridges is 
filamentary and decreases in surface brightness 
from NE towards the SW.  
The bright compact source that is observed in the image close to the NW
edge of the extended source, at RA $14^{h} 03^{m} 38\fs7$, 
DEC $-33\degr 58\arcmin 39\arcsec$
(J2000.0), has a flux density of 0.95 Jy at 330 MHz;
the compact source appears slightly resolved and has a 
deconvolved size of about 25 arcsec.  As discussed below, this compact
component is identified with the galaxy NGC 5419.
There is evidence for a protrusion towards the SW beyond the bounding
rim; this extension, marked P in Fig. 1, 
is observed in the 330 MHz image at a level of
$\sim 20$~mJy~beam$^{-1}$.
The total flux density of the extended radio source PKS~B1400$-$33
is 8.5 Jy at 330 MHz, excluding the emission from NGC 5419 and the 
source $5 \farcm 5$ to its south, marked B in Figs. 1-4, 
that is presumably an unrelated confusing source.  
The error in the absolute flux density scale at 330 MHz is believed
to be about 2 \%.  

The 843 MHz MOST image of PKS~B1400$-$33 is shown in Fig. 2. This image,
obtained using the most sensitive $23\arcmin$ field of view, 
better reproduces the extended structure than the
843 MHz image in \citet{goss87}.
The image has been made with a beam of FWHM $77 \times 43$ arcsec$^{2}$
at a P.A. of $0^{\circ}$ and has an r.m.s. noise of 1 mJy beam$^{-1}$.
The 843 MHz image of the extended source shows the relatively bright rim
along the NE, the filamentary structures trailing from this
ridge towards the SW and the curved rim that defines the boundary
of the diffuse source at the SW end.  
The flux density of the extended source (again excluding NGC 5419
and the embedded backgound source B) is 1.3 Jy at 843 MHz.  
NGC 5419 has a flux density of 0.46 Jy at 843 MHz.  
Additionally, the 843 MHz image clearly reveals a local emission peak 
embedded within the diffuse emission  
at RA $14^{h} 03^{m} 55\fs9$ DEC $-34\degr 06\arcmin 03\arcsec$ (J2000.0); 
we hereinafter refer to this feature as component C.  It has a flux 
density of 13 mJy~beam$^{-1}$ at 843 MHz.  The flux density scale
in the MOST image is accurate to 5 \%.

PKS~B1400$-$33 was observed using the 375 m and the 750B 750 m array 
configurations of the ATCA during 1995 Jan and June.  
Observations were made in two
128 MHz wide bands centred at 1398 and 2378 MHz; the bands were covered
using 13 channels.  The extended source
was covered with a nine-pointings mosaic with the pointings spaced 
$8\arcmin$ apart in RA and DEC.  Continuum images at the two frequencies were
made separately using the channel data, adopting a bandwidth synthesis approach
to avoid bandwidth smearing effects; the mosaic imaging was carried out
in {\sc aips++}.  Deconvolution used the multi-scale CLEAN algorithm:
the bright compact source at the edge of the extended emission was
first removed from the `dirty' image using a `box' CLEAN and subsequently
a multi-scale CLEAN was performed on the residual image. 
The final images at the two frequencies were then convolved to identical
beams of FWHM $70 \times 45$ arcsec$^2$ at a P.A. of $0^{\circ}$.
Consequently, different weightings were adopted at the two ATCA
frequencies during the imaging step; however, the convolution to
identical final beams implies that the images presented here effectively
have the same visibility range.
ATCA mosaic images of the PKS~B1400$-$33 field at 1398 and 2378 MHz are shown
in Figs. 3 and 4 respectively.  
The r.m.s. noise is 0.5 mJy beam$^{-1}$ in both images.
The relatively bright ridge along the NE and the filamentary
emission trailing towards SW have been imaged at 
1398; however, at 2378 MHz, the ridge appears non-uniform and clumpy 
and the filamentary emission is undetected,
possibly due to the poorer signal-to-noise ratio.
The bounding ridge at the SW of the extended source 
is not detected in either of these higher frequency ATCA images.
The wide-field mosaic
images show several continuum sources, presumably unrelated, in the field.
Based on the ATCA data, the flux density of NGC 5419 is
0.30 and 0.23 Jy at 1398 and 2378 MHz respectively.  The total flux density
of the extended source PKS~B1400$-$33 is 0.46
and 0.10 Jy at 1398 and 2378 MHz.  The central component C
is detected in both the ATCA images with flux density 5 and 2
mJy~beam$^{-1}$ at 1398 and 2378 MHz respectively.  The absolute flux density
scale in the ATCA observations was set using observations 
of PKS B1934$-$638 whose flux density is known, relative to
sources in the northern hemisphere, with an uncertainty of 2 \%.

The parameters of the radio images presented here are in
Table 2.  PKS~B1400$-$33 has an angular size 
of approximately $24\arcmin \times 14\arcmin$. Assuming
that the source is at the distance of the cluster S753, which is 
40$h^{-1}$ Mpc from the Sun ($h=H_{\circ}/100$,
where $H_{\circ}$ is the Hubble constant in km s$^{-1}$ Mpc$^{-1}$),
the linear dimensions of the extended source are 
approximately 280$h^{-1}$ kpc $\times$ 160$h^{-1}$ kpc.  
The 1.4 GHz radio luminosity
of the extended source is $2.2h^{-2} \times 10^{22}$ W Hz$^{-1}$.

\section{The distribution of the radio spectral index}

The radio spectra of the extended emission and of the compact source
associated with NGC 5419 are shown in Fig. 5.  
Apart from the measurements at 330, 843, 1398 and 2378 MHz reported here,
the plot includes previous estimates of the flux densities of the compact
and extended sources in \citet{goss87} as well as our estimates of the
85 and 408 MHz flux densities of the extended source that are based on the
measurements made by \citet{mills60} and \citet{bolton64} and extrapolations
of the flux density of NGC 5419 to low frequencies.

The extended radio source PKS~B1400$-$33 (not including NGC 5419 and the 
background source B)  has a 
mean spectral index $\alpha = 2.0$ between 330 and 1398 MHz. 
The spectral index
appears to steepen to $\alpha = 2.9$ between 1398 and 2378 MHz; this
may be because of missing extended flux density in the 2378 MHz 
interferometric mosaic image. The 85 MHz data indicates a spectral
flattening at low frequencies.
Our measurements of the flux density of the compact source associated 
with NGC 5419 are consistent with previous estimates; the compact
source has a spectral index $\alpha = 0.8$.  The compact source B located
$5\farcm5$ south of NGC 5419 
has $\alpha = 0.8$ between 843 and 1398 MHz
and $\alpha = 1.0$ between 1398 and 2378 MHz.

The images at 330 and 843 MHz
were convolved to a final beam of $80\arcsec$ FWHM and the distribution
of the spectral index, that was computed from these images, 
is shown in Fig. 6.  Effectively, the two images used for computing the
spectral index image have the same u,v-coverage.  Towards the NE ridge
$\alpha$ is 1.3-1.4 and towards the SW rim at the opposite
end of the diffuse source $\alpha$ is 1.6--1.7.   
In the region between the two rims, $\alpha$ is 1.7 towards
component C whereas the spectral index 
is steeper, with $\alpha$ in the range 1.8--2.4,
over most of the filamentary emission.

Between 1398 and 2378 MHz, 
the spectral index of the NE ridge is 1.9 and 
this is steeper than the spectral index between 330 and 843 MHz.   
Component C, which is detected in both the 1398 and 2378 MHz 
images, has a spectral index
of $\alpha = 1.7$ at these frequencies, similar to that 
between 330 and 843 MHz.

\section{The optical galaxy environment}

The 330 MHz radio contours are overlaid on a DSS digitization of
the UK Schmidt optical image of the field in Fig. 7.  
The bright compact radio source is coincident with the galaxy NGC 5419
which is the dominant galaxy of a cluster listed in the supplementary
catalog of \citet{abell89} as S753.  This is a poor cluster of Abell
richness class 0.  
\citet{willmer91} find that the mean harmonic
radius \citep{maia89} of the cluster members is 1.26$h^{-1}$ Mpc.
This radius corresponds to an angular size of $1\fdg8$. The extended source
PKS~B1400$-$33 is located well within the cluster radius; however, the
source is offset to one side of the central dominant galaxy.
Herein we assume that PKS~B1400$-$33 is at the distance of S753.
S753 has a high spiral content (45 \%) and a low velocity 
dispersion of 416 km s$^{-1}$ \citep{willmer91}.  Using different criteria
for cluster membership, \citet{fadda96} estimate the cluster internal 
velocity dispersion to be 536 km s$^{-1}$.

In the DSS optical image, there are no detectable optical 
counterparts that are 
positionally coincident with either the compact radio source $5\farcm5$ 
to the south of NGC 5419 or the central component C.

\section {The thermal gaseous environment}

The cluster is estimated to have a virial mass of 
$1.5 \times 10^{14}$ M$_{\odot}$ \citep{willmer91}. Based on the 
velocity dispersion - temperature relation given by \citet{bird95},
we estimate
that isothermal X-ray gas in the relatively low-depth potential well 
would have a temperature of 2--3 keV.  A 3 ks PSPC pointed observation
of the S753 cluster was recovered from the ROSAT archives.
We have examined the 0.1--2.4 keV broad-band 
image of the cluster.  There is a 
bright peak in the X-ray emission at the location of NGC 5419, which
might be emission associated with the interstellar medium of the central
galaxy or a result of a cooling flow.  There is a secondary peak, 
located about $3 \farcm 6$ to the west, that does not appear to
be associated with any cluster member. To detect any low surface brightness
diffuse X-ray emission (in the presence of the strong unresolved sources),
we convolved the $15\arcsec$ pixel counts on the sky image with a Gaussian of
FWHM $1\arcmin$ followed by a box-car of width $2\arcmin$. 
A contour representation of this X-ray image is shown in
Fig. 8 overlaid on greyscales of the 330 MHz VLA image.   
A low surface brightness extended X-ray halo 
component is observed, centered on NGC 5419, with
a radius of $16\arcmin$ (190$h^{-1}$ kpc).  The X-ray halo
appears to surround the radio source PKS~B1400$-$33. There is
no evidence in this ROSAT image for any deficit or excess of X-ray emission
towards the extended radio source.   The X-ray properties
of S753 are typical of poor groups \citep{mulchaey98}, where
the X-ray emission is dominated by a component associated with a bright
cluster member.

\section{The anomalously low surface brightness of PKS~B1400$-$33}

The physical parameters in those powerful
FR II radio galaxies and head-tail and wide-angle tail radio sources, which
have extended emission structures much bigger than the 
size of the host galaxy, may be useful probes of the surrounding 
medium \citep{feretti92,subrahmanyan93}.  
The radio structures that are overpressured with respect
to the ambient medium expand in the ambient gas with speeds
limited by ram pressure; therefore, derivations of the ambient gas
properties depend on a knowledge of the expansion speeds.  
Radio structures with relatively low surface brightness,
like the diffuse source PKS~B1400$-$33,
potentially have the lowest internal energy densities and pressures,
and are most likely to be in a static pressure balance with the ambient
thermal gas. Therefore, the internal state of such synchrotron 
plasma might be a direct probe of the surrounding thermal environment.

\citet{subrahmanyan93} studied the properties of the synchrotron plasma
in the diffuse bridges of the lobes of powerful and giant radio galaxies 
located in the field.  These structures, which lie outside cluster environments
and outside the ubiquitous thermal halos of the host ellipticals, 
are among the lowest surface brightness radio components outside clusters.  
These giant bridges have 1 GHz surface brightness 
$\sigma_{1~{\rm GHz}} \approx 0.2$ Jy arcmin$^{-2}$ \citep{subrahmanyan96} 
and the pressure inferred for the synchrotron plasma,
assuming minimum energy conditions \citep{miley80},
is $p_{e} = $1--2$\times 10^{-14}$ dyne cm$^{-2}$.  
A study of the internal pressures in the low surface brightness tails 
of tailed radio galaxies in Abell clusters \citep{feretti92} shows that
in cluster environments $p_{e}$ exceeds $10^{-13}$ dyne cm$^{-2}$.  
The higher pressures inferred for the radio galaxies in clusters 
are consistent
with the relatively higher ambient densities and temperatures in the
intracluster medium.  

Diffuse cluster radio sources, which cannot be identified with any
active radio galaxy, have a wide range in their surface brightness and 
$p_{e}$ in the range $10^{-14}$--$10^{-13}$ dyne cm$^{-2}$ \citep{feretti02}.
Among these sources, the peripheral relics, like the arcs in 
A3667 \citep{rottgering97}, have $\sigma_{1~{\rm GHz}}$ a factor 
10 larger than the giant bridges 
in the field and in this respect they are similar to the tailed radio
sources in cluster environments; however, the peripheral relics have been
inferred to have a wide range in their $p_{e}$ (see, 
for example \citet{feretti96}). The lowest surface brightness radio components 
in cluster environments are the diffuse cluster wide halos. Coma C,
a prototypical example of such a halo,
has $\sigma_{1~{\rm GHz}} \approx  3$ mJy arcmin$^{-2}$ \citep{kim90}
and an inferred $p_{e} \approx 8 \times 10^{-15}$ dyne cm$^{-2}$ 
\citep{giovannini93}.  Other cluster halo sources are inferrred to have
a similar low $p_{e} \la 10^{-14}$ dyne cm$^{-2}$ \citep{feretti96}. The 
extremely low synchrotron plasma pressure inferred for the halo sources appears
inconsistent with their location at the centers of rich clusters. This anomaly
might be related to the origin of the halo: the diffuse halos are not
identified with any currently active galaxy and are, instead, 
postulated as being reaccelerated relic plasma \citep{brunetti01}.  
The halos are believed to have an origin different from the peripheral relics.

The surface brightness of PKS~B1400$-$33
is about 3 mJy arcmin$^{-2}$ at 1 GHz.
If we assume that the cm-wavelength spectrum of PKS~B1400$-$33 continues
with $\alpha = 2.0$ to low frequencies, the inferred 
$p_{e}$ is $1 \times 10^{-13}$ dyne cm$^{-2}$. 
However, \citet{mills60} measured
the 85 MHz flux density of the source to be 57 Jy giving the source
a mean spectral index of $\alpha = 1.4$ between 330 and 85 MHz.
This indicates that the spectrum has a break somewhere in the range 
0.1--0.3 GHz and that the
spectral index flattens towards lower frequencies.  Assuming a
spectral break at 165 MHz with $\alpha = 0.7$ below the break and
$\alpha = 2.0$ above, we infer $p_{e} = 5 \times 10^{-14}$ dyne cm$^{-2}$. 

PKS~B1400$-$33 appears to be in a poor
cluster environment, where the ambient gas pressure 
is probably higher than that surrounding the bridges of giant radio galaxies 
in the field.  The cluster in which PKS~B1400$-$33
is located has a relatively low inferred gas temperature, a
low X-ray luminosity, and the ambient environment is not as extreme as 
that in rich clusters. 
PKS~B1400$-$33 has an extremely low surface brightness 
very similar to that of cluster wide halo sources such as Coma C.  
However, unlike the halo sources, PKS~B1400$-$33 is
not centrally located in the cluster.  Additionally, halos are almost
exclusively found in rich clusters with high velocity dispersions, whereas 
the cluster environment of PKS~B1400$-$33 is poor.  
The $p_{e}$ in PKS~B1400$-$33 is higher than that in giant bridges in the
field as well as diffuse halo sources in cluster centers, but
it is not as high as that in tailed radio
sources in rich clusters.  The intermediate value for $p_{e}$ is consistent
with PKS~B1400$-$33 being a relic in a poor cluster environment; however,
the surface brightness is relatively low as compared to typical relics.

Rood \#27 \citep{harris93} is another low surface brightness source 
in a poor galaxy environment.  This diffuse
source is not obviously identified with any optical galaxy and appears to
be a relic.  Rood \#27 also has an extremely low $\sigma_{1~{\rm GHz}}$
but somewhat higher than that of PKS~B1400$-$33; additionally, 
the spectral index of Rood \#27 is close to 0.6 and is not as steep 
as that of PKS~B1400$-$33.  

\section{On the nature of the unusual source PKS~B1400$-$33}

The extremely steep spectral index (mean $\alpha = 2$) of PKS~B1400$-$33
suggests that the source is composed of relic synchrotron plasma
in which energy injection has ceased and, subsequently, the spectral index 
has steepened as a consequence of synchrotron losses (spectral aging).
The steep spectrum indicates that there is no on-going or
recent reacceleration.  Moreover, the steep spectrum also suggests
that the emissivity of this source has not been significantly
enhanced as a result of adiabatic compression.  
PKS~B1400$-$33 is likely a relic of a source 
that was bright in its youth.  Extrapolating the measured flux density 
of 57 Jy at 85 MHz to cm wavelengths
using a spectral index of $\alpha=0.7$, we infer that the source would
have had a 1 GHz flux density at least 10 Jy prior to any losses.  The
radio power would have exceeded $4 \times 10^{24}$ W Hz$^{-1}$ at 1 GHz
and PKS~B1400$-$33 would have been a powerful radio source.

Our estimate that the diffuse source has a $p_{e}$ in the
range 5--10$ \times 10^{-14}$ dyne cm$^{-2}$ implies that the
equipartition magnetic field is 1-2 $\mu$G and spectral aging
is predominantly via inverse Compton losses against the cosmic
microwave background.
If we assume that the diffuse source was initially created radiating
with a power law spectrum and with spectral index $\alpha < 2$, 
and that the electrons with higher Lorentz factors were depleted 
as a consequence of spectral aging, the spectral break would move 
below 300 MHz in about $10^{8}$ yr.  If, today, the break is
at 165 MHz, the source age is $5 \times 10^{8}$ yr.

\subsection{A relic created by NGC 5419?}

PKS~B1400$-$33 has a radio power $0.9 \times 10^{23} h^{-2}$ W Hz$^{-1}$ 
at 1.4 GHz; before any spectral aging the radio power may have
been higher.  Luminous ellipticals with absolute magnitudes
$M_{R} \la -21$ are usually the hosts of extended 
radio sources with these high radio powers \citep{ledlow97}.  It follows
that the host galaxy of PKS~B1400$-$33 ought to be a bright elliptical
with $m_{R} \la 13.5$ (assuming $h=0.5$).  11 E/S0 galaxies listed
by \citet{willmer91} to be in S753
are of this magnitude and NGC 5419 is the most luminous.

NGC 5419 is a bright radio source 
suggesting that the central engine in the galactic nucleus is 
currently active.  However, based on Fig. 7, NGC 5419 is located
just outside the boundary of the extended source. Moreover, as seen
in Figs. 2 and 3, there is an emission gap between the compact source
associated with the galaxy and the extended source. Finally, the portion
of the extended source closest to the galaxy  has a different
spectral index ($\alpha = 1.6$) compared with the compact source
($\alpha = 0.8$).  Additionally, there is no evidence for
any spectral index gradient away from NGC 5419.  

Nevertheless, it might be that the extended source PKS~B1400$-$33 is a relic 
of past activity in NGC 5419.  In this case, the displacement of the 
extended structure from NGC 5419 could be due to 
Rayleigh Taylor instability of the light relic synchrotron plasma 
embedded in a denser X-ray gaseous environment and at the bottom of the
cluster potential well \citep{churazov00}.  Simulations of the rise
of such buoyant bubbles \citep{churazov01} indicate that the plasma
may initially deform into a torus as it rises and later form pancake-like
sheets at altitudes where the densities of the two-phase medium
attain equilibrium.  The arc-like boundaries of PKS~B1400$-$33 might be
fragments of such a global toroidal structure viewed in projection. 
In this scenario, the relatively flatter spectrum component C embedded
in the diffuse source might be the past site of nuclear activity.

The `ghost' cavities that are observed in the Perseus cluster \citep{fabian00}
and in Abell 2597 \citep{mcnamara01}, are believed to be relic synchrotron
bubbles. Buoyancy has been proposed as the mechanism for their displacement 
from the centers of the clusters.  If PKS~B1400$-$33 has been displaced from 
NGC 5419 over the distance of $r=100h^{-1}$ kpc by buoyant forces, we would
expect the movement to occur over a timescale 

\begin{equation}
t_{b} \approx \sqrt{(\rho_{e}/\rho_{ICM}) \times ( r^{3}/GM(<r)) },
\end{equation}

\noindent where $\rho_{e}$ is the density of the entrained thermal
matter, $\rho_{ICM}$ is the density of the thermal intra-cluster
medium and $M(<r)$ is the gravitational mass of the cluster within
radius $r$.  We infer that $M(<100h^{-1}~{\rm kpc}) \sim 7.5 \times
10^{10}$ M$_{\odot}$ from the cluster parameters derived by
\citet{willmer91} and, consequently, the timescale of the buoyant
motion $t_{b} < 1.6 \times 10^{9}$ yr. If we assume that the motion
takes place in the spectral aging timescale of $5 \times 10^{8}$ yr,
we infer that the ratio of ambient to entrained thermal material is
$(\rho_{ICM}/\rho_{e}) \approx 10$.

Alternatively, we consider the possibility that the displacement
between PKS~B1400$-$33 and NGC 5419 is the result of transverse motion
of the galaxy itself.  Assuming a timescale of $5\times 10^8$ years
and a separation $r=100h^{-1}$ kpc, the implied transverse velocity is
$190h^{-1}$ km s$^{-1}$.  This is consistent with the cluster velocity
dispersion and with the peculiar radial velocity of NGC 5419 of 184 km
s$^{-1}$ \citep{willmer91}.  The central component C, which has a
relatively flatter spectrum, might be the site where the galaxy was 
situated when activity in its nucleus created the extended source.

In the above scenario, PKS~B1400$-$33 was created in the past owing to 
nuclear activity in NGC 5419. Subsequently, the activity ceased, and
the current nuclear activity in NGC 5419 is a new activity phase.
The compact radio source associated with NGC 5419 
has a spectral index $\alpha =0.6$ that is flatter
than that observed in the extended emission; this gradient is
consistent with the hypothesis of restarting activity (see, for example,
\citet{roettiger94}).  If the spectral age derived
for PKS~B1400$-$33 is nearly the true age of the relic source, it follows
that the nuclear activity in NGC 5419 has restarted in 
less than $5 \times 10^{8}$ yr.  It may be noted that \citet{mcnamara01}
derived a timescale of $10^{8}$ yr for recurrent outbursts in the
central source of Abell 2597.

\subsection{Relic synchrotron plasma re-energized by shocks in the ICM?}

Clusters of galaxies are believed to be dynamically evolving at the
present epoch, accreting mass components and undergoing mergers, which
result in ICM gas density discontinuities that are either shock fronts
or cold fronts \citep{forman01}.  Relic radio sources
that have been rendered invisible at cm wavelengths 
owing to synchrotron losses might be
present in the intergalactic medium \citep{ensslin99}. 
\citet{ensslin98} have proposed that 
PKS~B1400$-$33 may be relic plasma in which the particles have been
reaccelerated by the passage of shocks related to the cluster evolution.
In this model, PKS~B1400$-$33 may have an origin external to S753.  

Steep spectrum relics are preferentially found at the peripheries of relatively
rich clusters usually with arc-like morphologies.  Examples are
the peripheral arc-like source J1324$-$3138 in A3556 \citep{venturi99} and 
the arcs in A3667 \citep{rottgering97}, located on two opposite sides 
of the cluster center with no detection of any diffuse extended emission
between. PKS~B1400$-$33 does indeed have edge-brightened arcs; however, 
in this case there is diffuse filamentary emission between and, 
additionally, the entire source is on one side of the cluster center.  

The morphological peculiarities in this relic might be a result of
an unusual projection geometry.  However, the parent cluster
S753 has a shallow gravitational potential well and the ICM presumably has
a relatively low gas pressure.  Consequently, relativistic electrons 
that are distributed over a wider area in the relic plasma may survive
radiative losses and be available to be re-energized by a 
passing shock wave \citep{ensslin01}.
The steep spectral index of PKS~B1400$-$33 (mean $\alpha = 2$) 
implies a relatively low shock compression ratio $R=2$;
this is consistent with the flatter 
gravitational potential of the poor cluster.
Simulations of the passage of a shock across a hot magnetized bubble 
\citep{ensslin01} suggest that the relic may transform into an
edge-brightened toroidal geometry.  The morphology of PKS~B1400$-$33 does 
suggest such an interpretation if the torus is being viewed almost
face on.

Extended halos are usually observed in clusters with a low spiral
fraction, large velocity dispersion and high X-ray luminosity.  These
are the clusters that are expected to have merger histories and large
scale shear and turbulence in the ICM and, consequently, halos.
However, S753 is a poor cluster and, although it belongs to the
Centaurus concentration of galaxies, it shows no evidence for
subclustering \citep{willmer91} or any anomalous distribution in the
velocities of its members which might be indications of ongoing or
past mergers.

As compared to other well studied halos and relic sources \citep{feretti02},
PKS~B1400$-$33 has a relatively small linear size and an extremely low
radio luminosity.  As discussed in section 6, the surface brightness 
of the source is extremely
low, similar to that in cluster halos; however, the inferred $p_{e}$
is more characteristic of cluster relics.  

\subsection{A relic of a lobe injected into the cluster environment?}

The radio galaxy phenomenon, in which beams from an active nucleus
power extended radio lobes, is believed to be short lived compared to
the Hubble time.  When the central engine switches off, presumably as the
fuel is exhausted, the lobes suffer radiative losses that 
deplete the more energetic electrons and steepen the spectrum.  However,
expansion losses, if present,  quickly render the synchrotron relic 
invisible because the radio power drops rapidly: in an expanding 
synchrotron bubble with a tangled magnetic field, 
the spectral luminosity falls 
as $L_{\nu} \sim f^{-(4\alpha +2)}$, where $f$ is the expansion 
factor \citep{leahy91}.  For this reason, we expect invisible relics
to be present in the intergalactic medium (IGM) where the ambient densities
and gas pressures may be lowest, and relics may preferentially be
visible in a higher density medium where expansion losses are smaller.  

A plausible scenario for the origin of a cluster
relic like PKS~B1400$-$33 would be one in which activity in a galaxy
located in the vicinity or boundaries of a cluster produced twin
beams, of which one resulted in a radio lobe within the ICM and the
other oppositely directed beam resulted in a second lobe outside of
the cluster gas and in a relatively lower density IGM.  The
differences between the environments of the two lobes would be more
pronounced if the double radio source is a giant radio source and the
activity axis is aligned with a local gradient in the ambient gas
density.  Following cessation of nuclear activity in the host, the
lobe in the IGM would quickly disappear from radio images, while 
the lobe located in the denser ICM would survive as a steep spectrum
relic.  

In this scenario, the extended source PKS~B1400$-$33 would be
one lobe of a relic double radio source.  From the viewpoint of such
an interpretation, the structure of the radio source (Fig. 1) suggests
that the double may have been edge brightened and the NE rim of
PKS~B1400$-$33 (the region with the flattest spectral index) might be
the site of past hotspots and the end of the double source.  The low
surface brightness filamentary emission trailing towards the SW from
this rim could be the relic cocoon of the backflowing plasma.  
The overall spectral steepening observed from the NE rim towards
the SW (see Fig. 6) is consistent with such a hypothesis.
The faint protrusion past the rim along the SW boundary (marked P in
Fig. 1) might be a relic
bridge.  We would expect a possible host galaxy to lie towards the SW and
along the axis defined by the protrusion; a counterlobe, if it is
detectable, may be located beyond that.

To test this hypothesis, we first searched wide-field NVSS and MOST
images of this field for a possible counterlobe along the axis defined
by the brightest portion of the NE arc and the SW protrusion in
Fig. 1. A tentative MOST detection at 843 MHz was followed up with a
pointed observation using the $23\arcmin$ field of view.  When
smoothed to a resolution of $2\arcmin$ FWHM, this observation gives
marginal evidence for a very extended source ($\sim$50 arcmin$^2$)
with centroid position RA $13^{h}59^{m}30^{s}$, DEC
$-34\degr47\arcmin$ (J2000.0) and an integrated flux density of $50
\pm 12$ mJy.  We then examined the optical field along the same axis
for possible host galaxies.  A candidate with $m_B=16.7$, and
classified as S0 by \citet{willmer91}, was found at RA
$14^{h}02^{m}18\fs3$, DEC $-34\degr22\arcmin54\arcsec$ (J2000.0); it 
is marked H in Fig. 7. The peak of PKS~B1400$-$33, the location of
galaxy H and the peak of the extended source tentatively detected in the
MOST observation are collinear within a few degrees.

We subsequently obtained an optical spectrum of galaxy H in the
4000--7700 \AA\ window using the ANU 2.3-m telescope and double-beam
spectrograph at Siding Spring Observatory in May 2002.  Its redshift,
$z=0.01695 \pm 0.0002$, implies a peculiar velocity of 1090 km
s$^{-1}$ relative to the mean redshift of the cluster.  Examining the
distribution in galaxy velocities for a magnitude limited sample
towards the cluster (see Fig. 4 in \citet{willmer91}), we infer that
galaxy H is likely to be a member of S753.  It lies $26\arcmin$ from
the center of PKS~B1400$-$33, implying a projected separation of
300$h^{-1}$ kpc at the distance of the cluster. The optical spectrum
shows narrow emission lines of H$\alpha$, [N II] and the [S II]
doublet; the [N~II]/H$\alpha$ and [S II]/H$\alpha$ ratios are about
0.25 and 0.22 strongly suggestive of starburst activity (and not
emission from an AGN environment).

AGN-type radio activity usually occurs in elliptical galaxies and the
probability of such activity is an increasing function of optical
luminosity.  Samples of extended radio sources show a sharp decline
for hosts with $R$-band optical luminosity $M_{R} \ga -21$
\citep{ledlow97}.  On the other hand, in the cluster Abell 428,
\citet{ledlow98} do find a powerful extended radio source 0313$-$192
associated with an $M_{B} = -19.9$ disk-dominated host that is most
likely an early type spiral (Sa--Sb).  The argument against galaxy H
being the host of PKS~B1400$-$33 is simply that it is even further
underluminous, with $M_{B}=-17.4$.

\section{Summary}

We have presented new and improved radio images of the relic source 
PKS~B1400$-$33.  We have discussed the origin of this relic in the light 
of recent progress in the understanding of relic synchrotron plasma in 
cluster environments.  The diffuse source has an extremely low surface 
brightness and a steep spectrum, with radio properties unlike 
that of relics and halos observed in cluster environments.  The unusual
morphology, placement in the cluster and physical parameters indicate
an unusual origin for this source.  

The new data does not represent
evidence suggesting that the relic source PKS~B1400$-$33 was created 
by past activity in NGC 5419.  Nor do the observations rule out this 
possibility.  The discussions suggest that PKS~B1400$-$33 is not a 
typical relic.  The radio properties suggest
that if the source was created by processes similar to those that
form the relics and halos, we might be observing an extreme form of
a relic in a poor cluster environment or, perhaps, a relic of a 
re-energized relic.  We have presented marginal evidence 
supportive of the hypothesis
that the diffuse source PKS~B1400$-$33 was born as a lobe of a powerful
radio galaxy; however, we do not have a good candidate for the
optical host.  The source is unusual and the peculiar radio properties
might be indicating an unusual origin: we favour the interpretation
that the diffuse source is a relic of a lobe
injected into the cluster environment as worthy of 
further consideration, particularly
because of the novelty of the proposed phenomenon. Followup low frequency
imaging of the field, possibly with the VLA at 74 MHz, with the aim of
confirming the MOST detection of a counterlobe and better quality imaging 
of any low surface brightness connecting features, is proposed as
the next step towards understanding the phenomenology in this
unusual source.

\acknowledgments

The Australia Telescope Compact Array is part of the Australia
telescope which is funded by the Commonwealth of Australia for
operation as a National Facility managed by CSIRO. The National
Radio Astronomy Observatory is a 
facility of the National Science Foundation operated under cooperative
agreement by Associated Universities, Inc.

\clearpage

\begin{deluxetable}{lcll}
\tablewidth{0pt}
\tablecolumns{4}
\tablecaption{Journal of observations. \label{tab1}}
\tablehead{
\colhead{Frequency (MHz)} & \colhead{Telescope} & 
\colhead{Epoch} & \colhead{Configuration}}
\startdata
1398 and 2378 & ATCA & 1995 Jan & 375m array \\
& & 1995 June & 750B array \\
843 & MOST & 1998 May & $23\arcmin$ f.o.v. \\
330 & VLA & 2000 March & CnB array \\
& & 2000 May, June & C array \\
& & 2000 July  & DnC array \\
\enddata
\end{deluxetable}

\clearpage

\begin{table}
\caption{Image parameters and total flux density of 
the extended source \label{tab2}}
\begin{tabular}{cccc}
\\
\tableline
\tableline
\\
Frequency & Beam FWHM & Image r.m.s. noise & Flux density \\
(MHz) & & (mJy beam$^{-1}$) & (Jy) \\
\\
\tableline
\\
330   & $77\arcsec \times 53\arcsec$ at P.A. $-24\degr$ & 2.5 & 8.5 \\
843   & $77\arcsec \times 43\arcsec$ at P.A. $0\degr$ & 1.0 & 1.3 \\
1398  & $70\arcsec \times 45\arcsec$ at P.A. $0\degr$ & 0.5 & 0.46 \\
2378  & $70\arcsec \times 45\arcsec$ at P.A. $0\degr$ & 0.5 & 0.10 \\
\\
\tableline
\end{tabular}
\end{table}

\clearpage

\begin{figure}
\epsscale{1.0}
\plotone{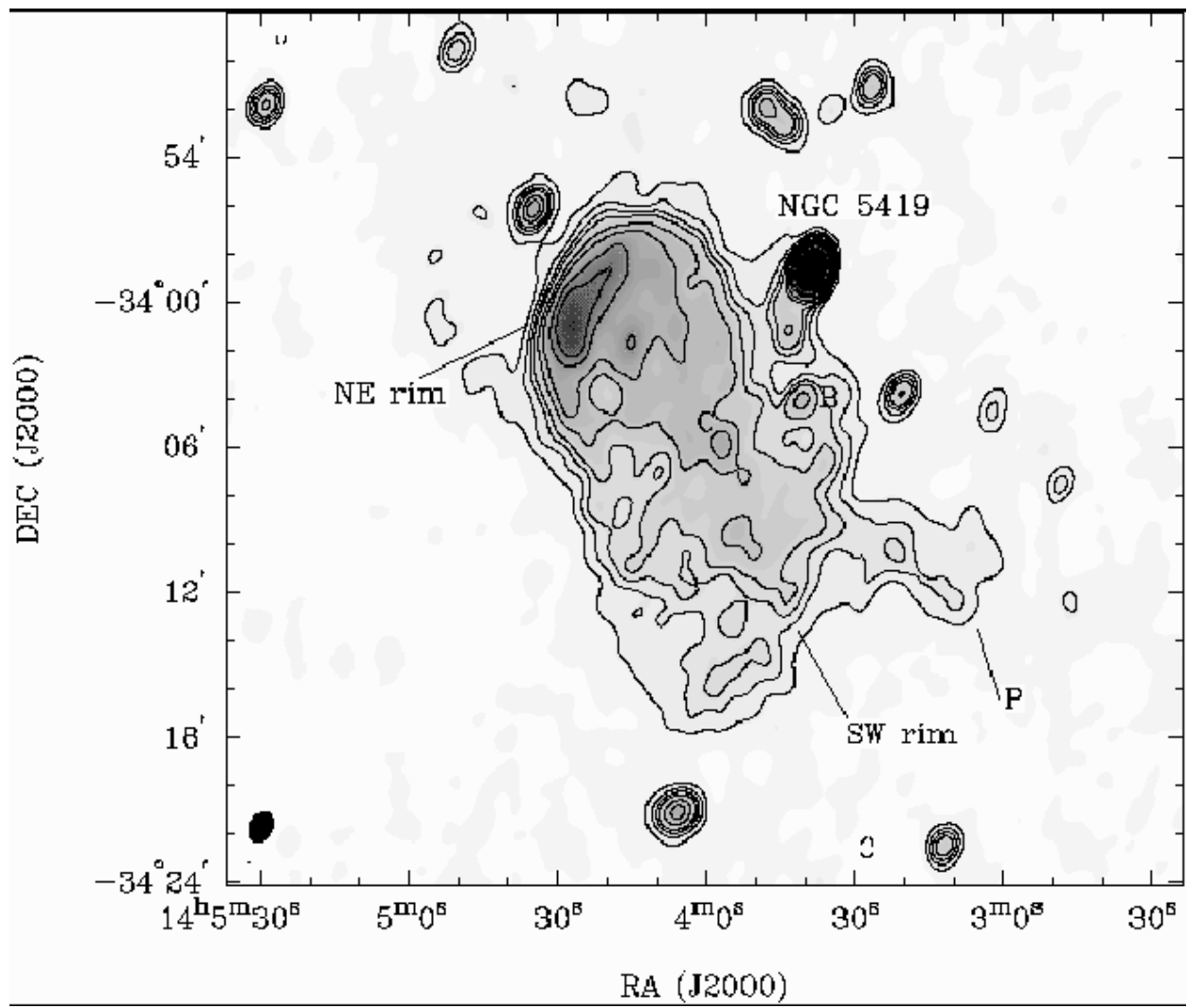}
\caption{PKS~B1400$-$33 at 330 MHz shown as a greyscale image with 
contours overlaid.  The VLA image has been made with a 
beam of FWHM $77\arcsec \times 53\arcsec$ at a 
P.A. of $-24\degr$; contours are at 10 mJy~beam$^{-1} \times(-1$, 1,
2, 3, 4, 6, 8, 12, 16, 24, 32, 48, 64, 96, 128). The r.m.s. noise is about
2.5 mJy beam$^{-1}$ in source free regions.  The NE rim, SW rim and protrusion
towards the SW, which have been referred to in the text, 
have been labelled in the figure.
This image, as well as all others displayed in this paper,
has been corrected for the attenuation due to the primary beam;
the shaded ellipses in the lower left corner of the images show the
half-power size of the synthesized beams. \label{fig1}
	}
\end{figure}

\clearpage

\begin{figure}
\epsscale{0.9}
\plotone{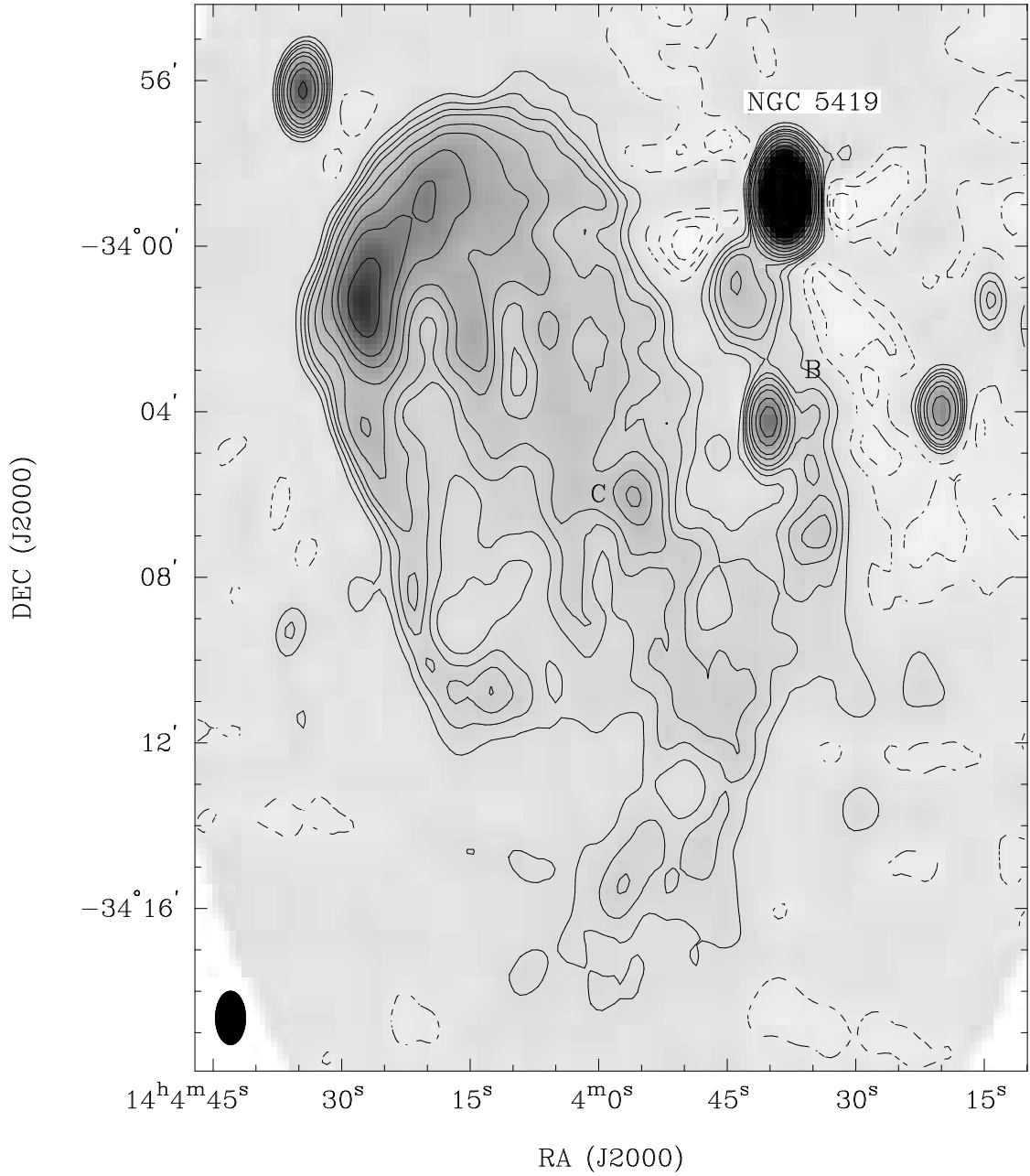}
\caption{ 843 MHz MOST image of PKS~B1400$-$33 made with a beam of 
FWHM $77\arcsec \times 43\arcsec$ at a 
P.A. of $0\degr$. The image noise is 1 mJy beam$^{-1}$.   
Contours are at 2 mJy~beam$^{-1} \times(-3$, $-$2, 
$-1$, 1, 2, 3, 4, 6, 8, 12, 24, 32, 48, 64, 96, 128).
 \label{fig2}
	}
\end{figure}

\clearpage

\begin{figure}
\epsscale{0.9}
\plotone{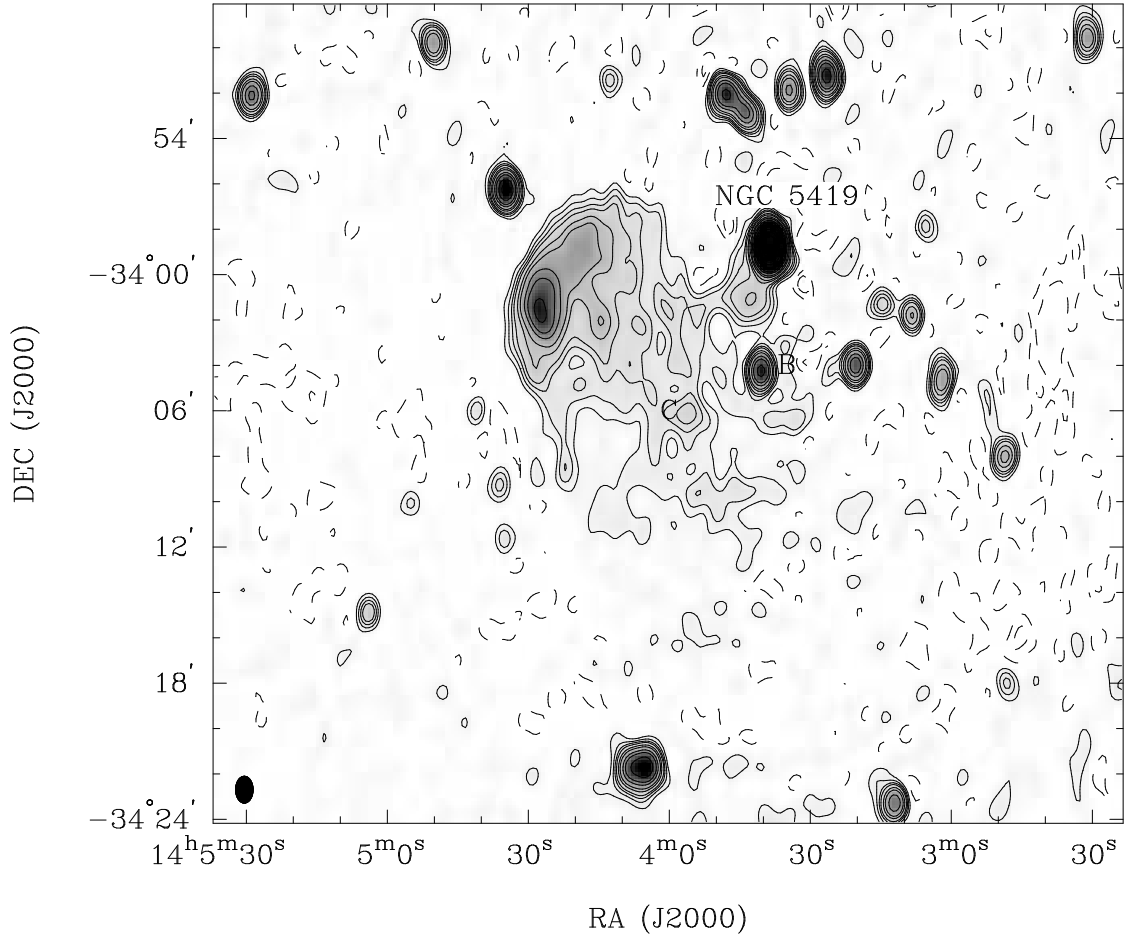}
\caption{ ATCA image of PKS~B1400$-$33 made at 1398 MHz with a beam of 
FWHM $70\arcsec \times 45\arcsec$ at a P.A. of $0\degr$.  The field
was mosaic imaged with 9 pointing centers spaced $8\arcmin$ in RA
and DEC.  The image
r.m.s. noise is 0.5 mJy beam$^{-1}$;  contours are 
at $-3$, $-$2, $-1$, 1, 2, 3, 4, 6, 8, 12, 24, 32, 48, 64, 96, 
128, and 192 mJy beam$^{-1}$.
 \label{fig3}
	}
\end{figure}

\clearpage

\begin{figure}
\epsscale{0.9}
\plotone{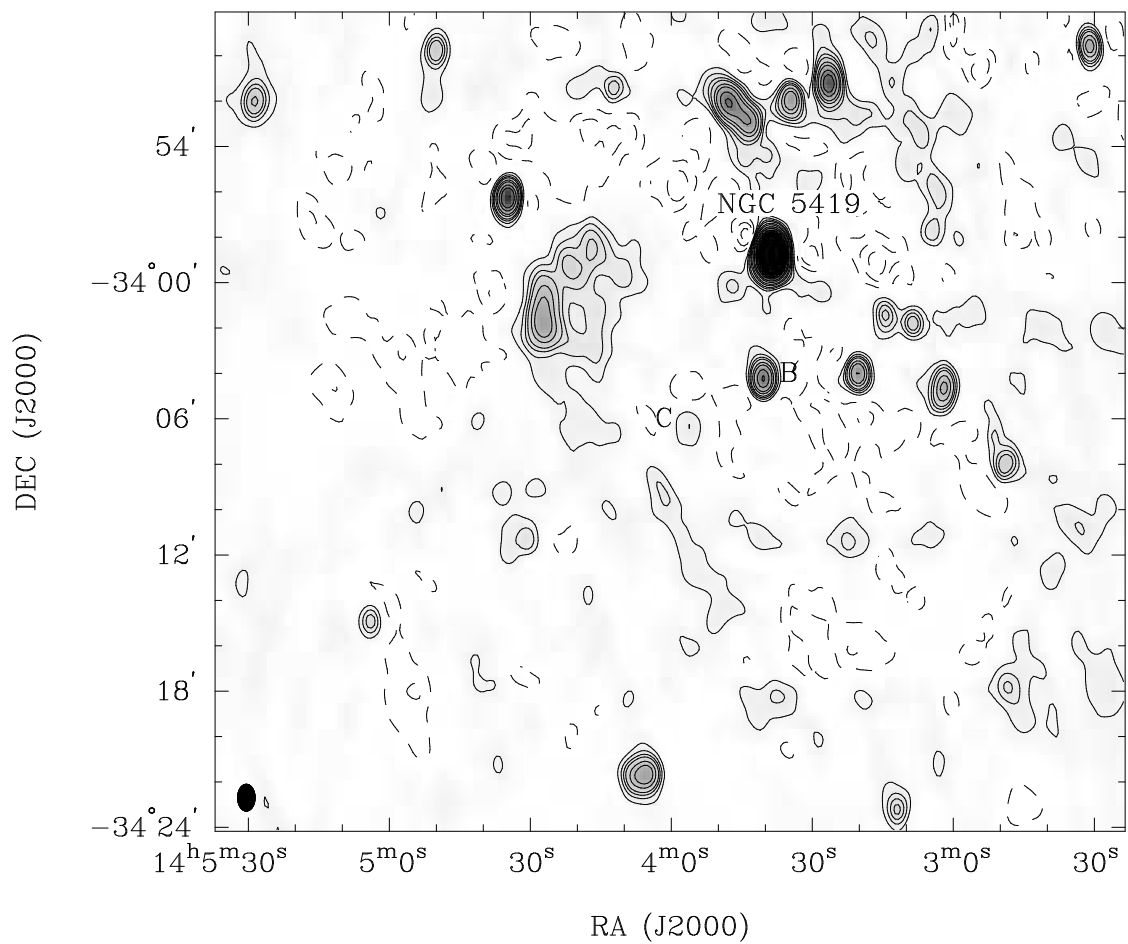}
\caption{ ATCA image of PKS~B1400$-$33 made at 2378 MHz with a beam of 
FWHM $70\arcsec \times 45\arcsec$ at a P.A. of $0\degr$. The observations
were made as a 9 pointing mosaic with fields spaced by $8\arcmin$
in RA and DEC.  The image
has an r.m.s. noise of 0.5 mJy beam$^{-1}$; contours are 
at $-3$, $-$2, $-1$, 1, 2, 3, 4, 6, 8, 12, 24, 32, 48, 64, 96, 
128, and 192 mJy beam$^{-1}$.
 \label{fig4}
	}
\end{figure}

\clearpage

\begin{figure}
\epsscale{1.0}
\plotone{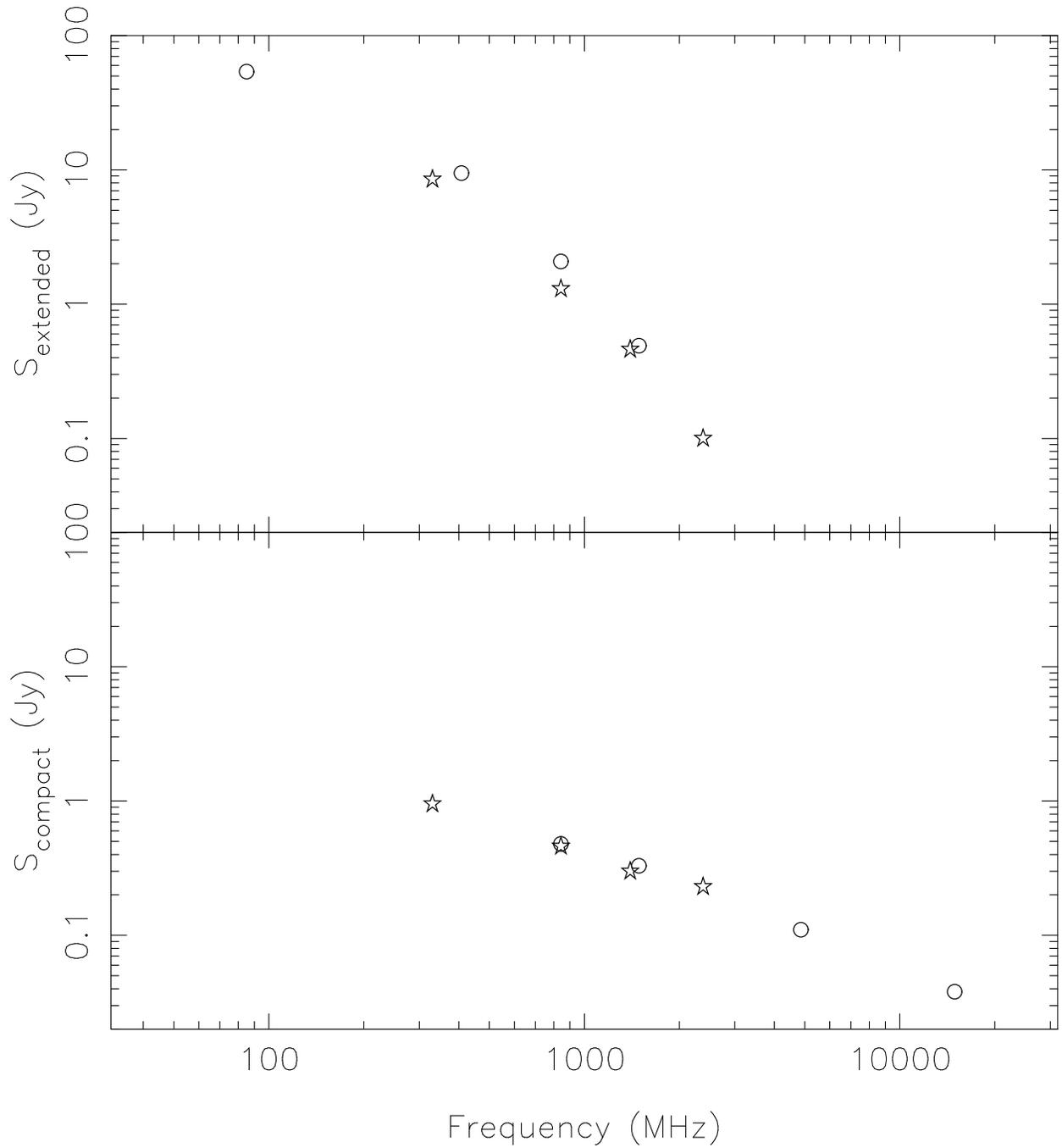}
\caption{ Radio spectra of the extended emission (upper panel) and
the compact source associated with NGC 5419 (lower panel).  The 
open circles represent previously published measurements, 
while the star symbols show flux density
measurements derived from observations presented in this paper. 
 \label{fig5}
	}
\end{figure}

\clearpage

\begin{figure}
\epsscale{0.9}
\plotone{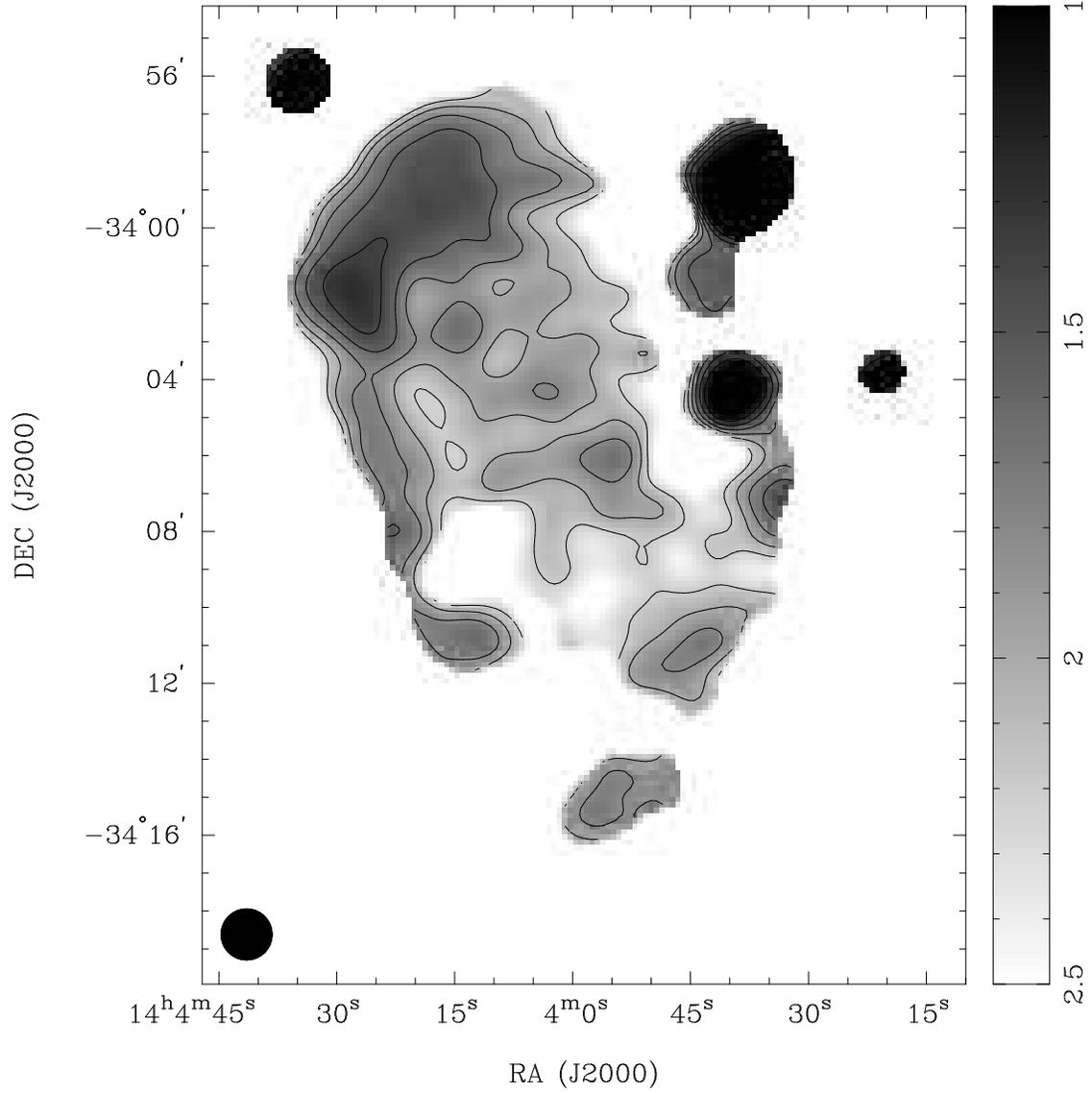}
\caption{ Distribution of the spectral index $\alpha$ 
between 330 and 843 MHz as
computed from images made with a beam of FWHM $80\arcsec$. The $1-\sigma$
uncertainty is $\pm 0.06$ in $\alpha$. Contours are at $\alpha$ values of
2.2, 2.0, 1.8, 1.6, 1.4, 1.2 and 1.0.
 \label{fig6}
	}
\end{figure}

\clearpage
\begin{figure}
\epsscale{1.0}
\plotone{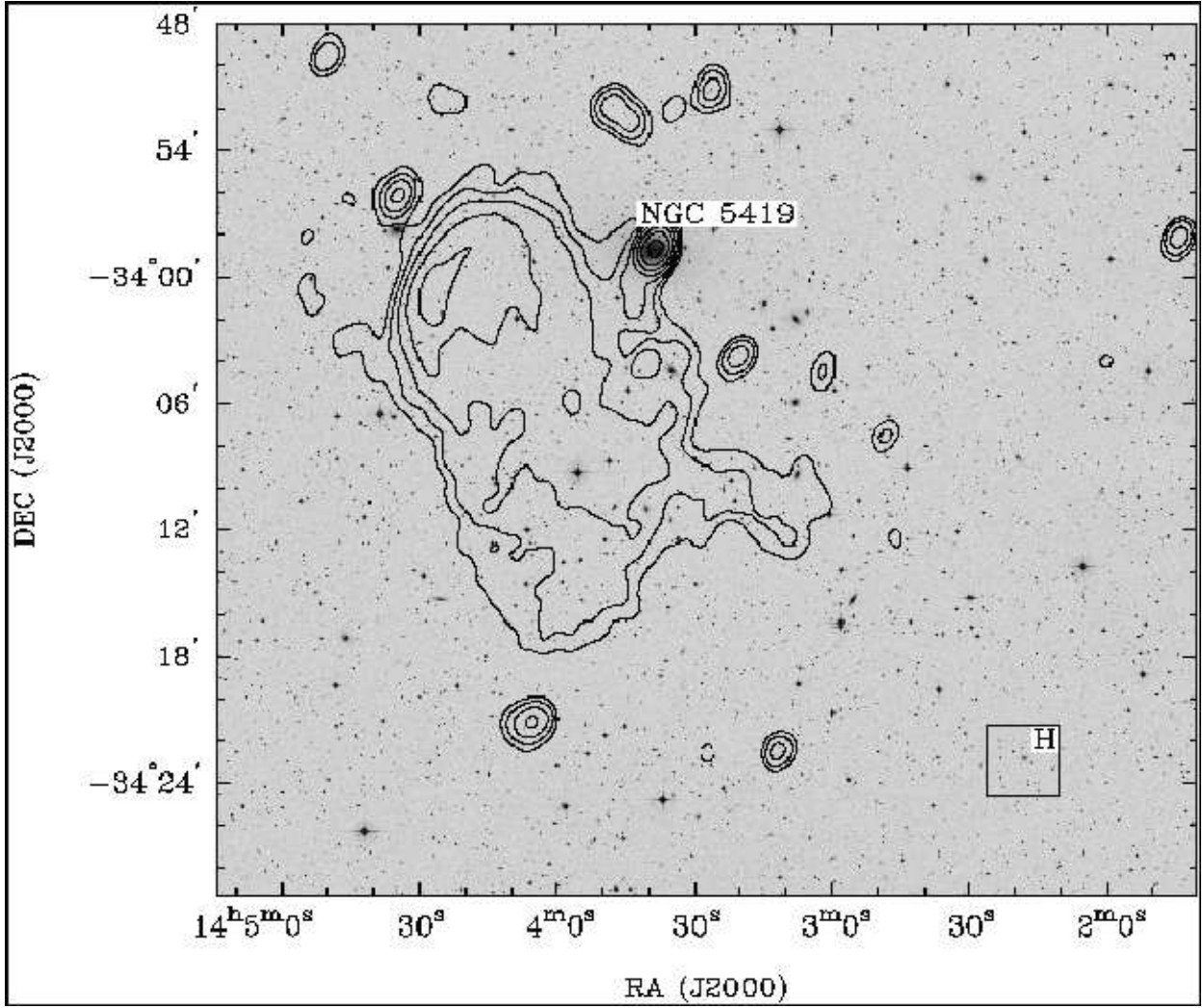}
\caption{ 330 MHz radio contours overlaid on a DSS digitization of the
UK Schmidt SERC-J survey image of the field.  Contours are at 
10 mJy~beam$^{-1} \times(-1$, 1, 2, 4, 8, 16, 32, 64). The compact radio
source identified with NGC 5419 is indicated; the candidate host galaxy
discussed in section 7.3 is at the center of the box marked H.
 \label{fig7}
	}
\end{figure}

\clearpage

\begin{figure}
\epsscale{1.0}
\plotone{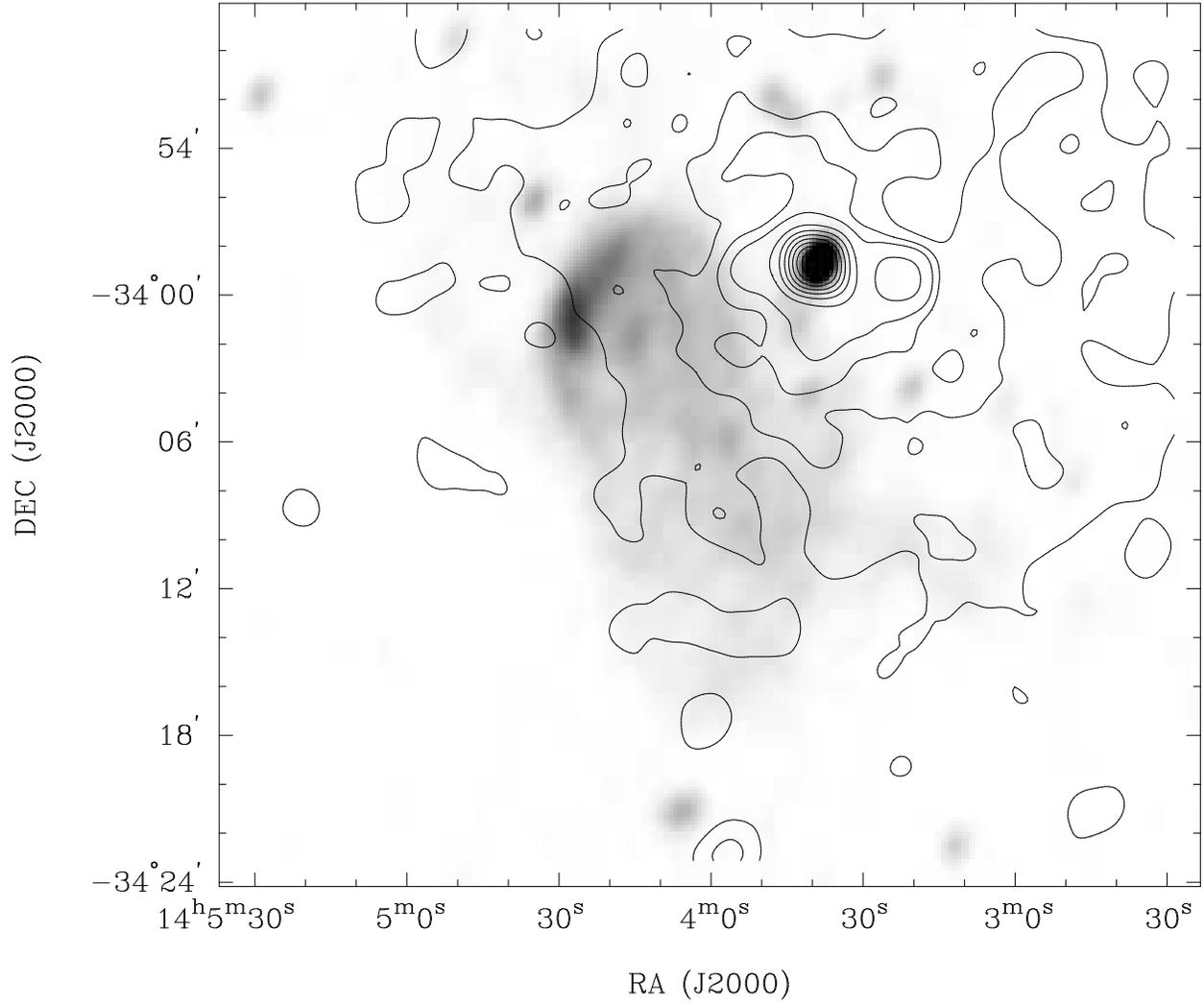}
\caption{ Contours of the ROSAT PSPC broad band X-ray image 
of the S753 cluster field overlaid on a greyscale representation
of the 330 MHz VLA image shown in Fig. 1. The X-ray image has been 
smoothed to a resolution of $2\farcm2$ FWHM and the contours are at 
8, 12, 16, 20, 30, 40, 50, 60, 70, 80 and 90 \% of the peak.
 \label{fig8}
	}
\end{figure}

\end{document}